# Phase-matched high-order harmonic generation in pre-ionized noble gases

O. Finke,[1,2] J. Vábek,[1,2,3] M. Nevrkla,[1,2] N. Bobrova,[2] O. Hort,[1] M. Jurkovič,[1,2] M. Albrecht,[1,2] A. Jančárek,[1,2] F. Catoire,[3] S. Skupin,[4] and J. Nejdl[1,2,*]

[1]*ELI Beamlines Centre, FZU- Institute of Physics of the Czech Academy of Sciences, Na Slovance 2, 182 21 Prague, Czechia*
[2]*Faculty of Nuclear Sciences and Physical Engineering CTU, Břehová 7, 115 19 Prague 1, Czechia*
[3]*Centre des Lasers Intenses et Applications, Université de Bordeaux-CNRS-CEA, 33405 Talence Cedex, France*
[4]*Institut Lumière Matière, UMR 5306 Université Lyon 1 - CNRS, Université de Lyon, 69622 Villeurbanne, France*
*\*nejdl@fzu.cz*

**Abstract:** One of the main difficulties to efficiently generating high-order harmonics in long neutral-gas targets is to reach the phase-matching conditions. The issue is that the medium cannot be sufficiently ionized by the driving laser due to plasma defocusing. We propose a method to improve the phase-matching by pre-ionizing the gas using a weak capillary discharge. We have demonstrated this mechanism, for the first time, in absorption-limited XUV generation by an 800 nm femtosecond laser in argon and krypton. The ability to control phase-mismatch is confirmed by an analytical model and numerical simulations of the complete generation process. Our method allows increasing the efficiency of the harmonic generation significantly, paving the way towards photon-hungry applications of these compact short-wavelength sources.

## I. INTRODUCTION

High harmonic generation (HHG) in rare gases is a compact tabletop source of coherent extreme ultraviolet (XUV) radiation used in a large variety of research fields [1-3]. A very successful semiclassical microscopic description developed almost 30 years ago [4] states that incoming laser radiation disrupts the atomic barrier potential, leading to ionization. The freed electron quivers along the linearly polarized laser field and recombines later with the parent ion within a half-cycle time period, emitting the excess energy as XUV radiation. Besides the effects influencing this microscopic phenomenon, the overall XUV signal strength is mainly affected by the phase-matching between the driving IR laser pulse and the generated high harmonic radiation [5,6].

With the increasing peak power of laser systems currently available for driving HHG [7,8], the f-number and the length of the generating medium increase following scaling laws [9,10] to maintain the maximum conversion efficiency with a uniform generating medium [11]. Several approaches to control the phase-matching were examined, such as modifying the wavefront of the driving laser field [12-14] or changing the properties of the generating medium by mixing multiple rare gases [15]. Other studies were devoted to guiding the beam and generating harmonics in a highly ionized medium [16-18]. Not many studies, however, have examined the impact of nonlinear laser pulse propagation on phase-matching in long medium.

In this article, we first show that only limited phase-matching is reachable in a long medium of neutral noble gas because of ionization-induced defocusing of the laser pulse. Then we introduce a controlled low pre-ionization of the medium as a method to overcome this bottleneck and verify it experimentally for what we believe is the first time. Additionally, an analytical model for defining the phase-matching conditions in a pre-ionized medium is derived to support the experimental results obtained with a weak capillary discharge. Moreover, we have performed numerical simulations of the whole process, modeling both microscopic and macroscopic aspects of HHG, to validate the method in a broader range of experimental conditions.

## II. PHASE-MATCHING LIMITATIONS DUE TO DEFOCUSING IN LONG MEDIUM

The impact of ionization-induced defocusing on the shape of a laser beam in long media has been examined. It was shown that there is a drastic drop of the laser intensity along the direction of propagation once the entry intensity is higher than a critical value [19,20]. To illustrate the importance of such behavior in more detail, we have performed numerical simulations of an 800-nm laser pulse propagation in a 15 mm long gaseous krypton medium for different laser intensities and gas pressures. (Fig. 1a, see section 5 for details about the numerical model). Similar to [21], the results of our simulations show that the intensity drop (from 0 to 6mm) is followed by a stable region with almost constant laser intensity. The value of stabilized intensity depends only on the medium type and density, and is practically independent of the laser intensity at the entrance of the medium. We can thus recognize two regions of generation in the long media: region I, where the pulse intensity is rapidly dropping during propagation and region II, the stable region.

Considering HHG in a long medium, the XUV reabsorption plays a critical role as soon as the absorption length of the generated harmonic is shorter than the total length of the medium. This is typically the case when generating harmonics below 100 eV with an 800 nm laser [22]. Therefore, achieving phase-matching in region II is crucial, as most of the detected XUV signal originates from this region.

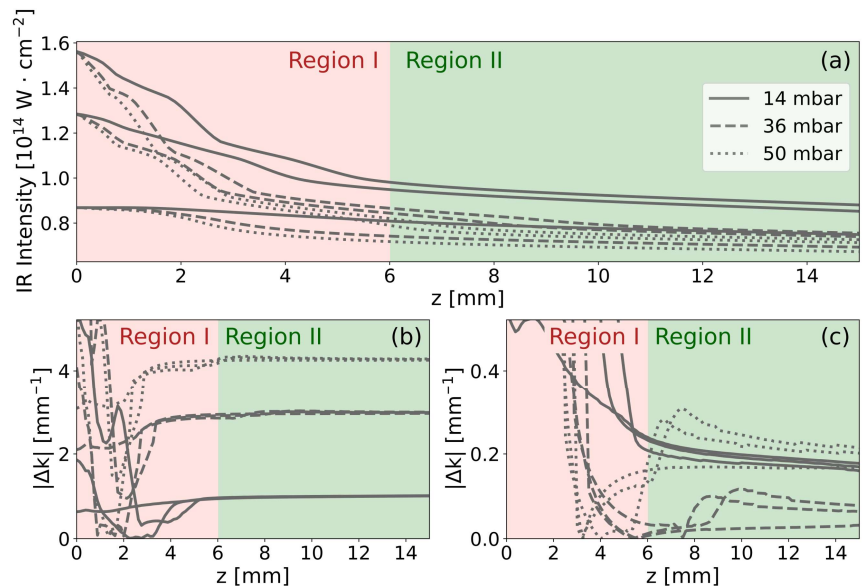

Fig. 1. (a) Results of numerical simulations of a 30 fs pulse propagating in krypton demonstrating the stabilization of on-axis intensity for different medium pressures and input laser intensities with emphasis on region I in light red, where intensity drops rapidly and region II in light green, where intensity stabilizes (b) Evolution of phase-mismatch $\Delta k$ between the 800-nm laser and the 17th harmonic order calculated for the same pulse propagation as panel (a). (c) Phase-mismatch for the same laser pulse as in (b), but with the medium homogeneously pre-ionized to 8 %. The medium length is set to 15 mm.

However, the intensity in region II is generally insufficient to ionize the medium by the driving laser itself to allow reaching the phase-matching condition. The central optical quantity to classify the phase-matching of two waves is their phase mismatch

$$\Delta k = k_q - q k_1, \qquad (1)$$

where $k_q$ and $k_1$ are the wavenumbers of the $q$-th harmonic and the fundamental, respectively. Ideal phase-matching corresponds to a situation where $|\Delta k| = 0$. Figures 1b and 1c compare the mismatch for the 17$^{\text{th}}$ harmonic and 800 nm laser from a numerical experiment with and without the pre-ionization (see the Supplement for detailed analysis). It demonstrates a significant reduction of the mismatch by homogeneously pre-ionizing the medium to ionization degree of 8 %. In this case, the $|\Delta k|$ is decreased by an order of magnitude in region II where an almost perfect phase-matching is achieved (Fig. 1c, note the different y-scale in Fig. 1b and 1c).

### III.  ANALYTICAL MODEL

We now derive an analytical model that allows us to quantify the optimal pre-ionization of the medium and the XUV signal enhancement.

A macroscopic model of HHG in a medium of length $L_{med}$ with the absorption of generated radiation characterized by absorption length $L_{abs}$ was developed earlier [23] providing an analytical estimate of the HHG gain in a homogeneous medium as

$$I_q = \frac{4\rho^2 A_q^2 L_{abs}^2}{1 + 4\pi^2 \left(\frac{L_{abs}}{L_{coh}}\right)^2} \left[1 + \exp\left(-\frac{L_{med}}{L_{abs}}\right) - 2 \cos\left(\frac{\pi L_{med}}{L_{co}}\right) \exp\left(-\frac{L_{med}}{2 L_{abs}}\right)\right], \qquad (2)$$

where $A_q$ is the dipole amplitude for the $q$-th harmonic (considered as constant in this model), $\rho$ is the medium density, and $L_{co} = \pi/|\Delta k|$ is the coherence length, i.e. the distance along which all microscopic emitters add constructively. In the case of absorption-limited generation ($L_{med} \gg L_{abs}$), which is usually the condition of generation, the pre-factor in (2) dominates as the expression in square brackets approaches unity.

If the phase matching is achieved and the coherence length becomes indefinitely large, the denominator of the pre-factor decreases to unity, and we get the phase-matched absorption-limited gain which is independent of medium density as $L_{abs} = 1/\sigma\rho$, with $\sigma$ being the absorption cross-section. With definite coherence length, it is the ratio of $L_{coh}$ and $L_{abs}$ that determines the gain of absorption-limited HHG. More generally, as the coherence length is inversely proportional to density, the absorption-limited gain becomes pressure independent.

The coherence length $L_{coh}$ is defined by the mismatch of the wavenumbers of the driving laser $k_1$, and the generated q-th harmonic beam $k_q$. We consider following dominant processes for each contribution to the wavenumbers: the dispersion of neutrals for both fields [24-26] and the effect of plasma on the medium susceptibility for the fundamental laser field. Effects of dipole phase on the harmonic generation and the Gouy phase of the Gaussian beam are neglected. Using susceptibilities in the expression for refractive indices of neutrals and plasma, we get

$$\Delta k = \frac{q\omega_1 \rho}{2c}\left(\Delta\chi - \eta \frac{e^2}{\varepsilon_0 m_e \omega_1^2}\right), \quad (3)$$

where $\Delta\chi = \chi_{1n}^{(a)} - \chi_{qn}^{(a)}$ is the difference of electric susceptibilities (at standard temperature and pressure) and $\eta$ is the degree of ionization, i.e., the ratio of free electrons to neutral atoms and ions. Inserting this expression in the definition of the coherence length, we can express the degree of ionization $\eta$ as a function of atomic density and coherence length:

$$\eta = \frac{\omega_1^2 \varepsilon_0 m_e}{e^2}\left(\Delta\chi \mp \frac{2\pi c}{q\omega_1 \rho L_{coh}}\right). \quad (4)$$

We consider ideal phase-matching as a situation where coherent length is comparable or longer than the medium length. This is the case for a range of degrees of ionization. Figure 2a shows the boundaries of $\eta$ as a function of pressure with coherence length equal to medium length for the 17th harmonic order generated by 800 nm laser in krypton. The case corresponding to an infinite coherence length is shown by the dashed line. As the laser ionizes the medium during generation, HHG is phase-matched as long as the ionization stays within the region shown in blue. This model explains the numerical observation presented in Fig. 1b and Fig. 1c. where the phase mismatch is strongly reduced in the region II when the medium is pre-ionized.

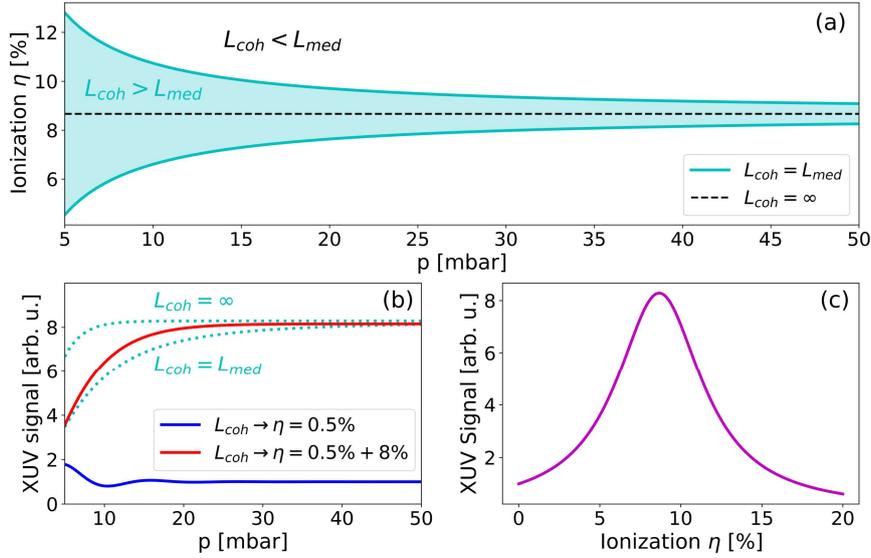

Fig. 2. (a) Range of optimal ionization ratio with $L_{med} = 15\ mm$ for 17th harmonic order generated by 800-nm laser in krypton as a function of pressure (cyan channel) with emphasis on ideal ionization corresponding to $L_{coh} \to +\infty$ (black dashed line). (b) Describes evolution of the 17th harmonic signal based on (2), where optimal signal is in range $L_{co} \in \langle L_{med}, +\infty \rangle$ (cyan dotted curves). Expected signal from the high-intensity laser in region II shown in Fig 1b is denoted by blue curve. By adding constant pre-ionization of 8% to the same high-intensity laser, the expected signal moves into desired region (red curve). (c) The absorption-limited XUV signal as function of ionization ratio $\eta$ according to (4).

Using the expression of the absorption and coherence lengths and inserting them into the pre-factor of (2), we obtain the absorption-limited signal

$$I_q(\eta) = \frac{4A_q^2}{\sigma^2 + \frac{q^2\omega_1^2}{c^2}\left(\Delta\chi - \eta \frac{e^2}{\omega_1^2 \varepsilon_0 m_e}\right)^2}. \quad (5)$$

The most striking result is that the only variable is the degree of ionization $\eta$. In particular, this expression is independent of the medium density. And it is also independent on the medium length (consequence of the assumption $L_{med} \gg L_{abs}$). The signal strength of 17th harmonic order generated in Kr given by (5) is shown in

Fig 2c as a function of degree of ionization $\eta$. The maximum of the signal is reached for $\eta_{\text{opt}} = \Delta\chi \frac{\omega_0^2 \varepsilon_0 m_e}{e^2}$. The possible enhancement, $I_q(\eta_{\text{opt}})/I_q(0)$, turns out to be 8 for the 17$^{\text{th}}$ harmonic in krypton. Note that $I_q(\eta_{\text{opt}})/I_q(0)$ depends only on material constants and laser frequency, it thus provides a direct way to estimate possible enhancement by the pre-ionization in absorption limited regime.

The effect of pre-ionization is demonstrated in figure 2b, where pressure dependence of the 17$^{\text{th}}$ harmonic order signal generated in a 15-mm long krypton medium is calculated from (2). XUV signal for the ideal generation is expected in the region between $L_{coh} = L_{med}$ and $L_{coh} = \infty$. These curves represent the best possible signals and show, that pressure up to $\sim$35 mbar is generally desired. We took the intensity from the region II of Fig. 1b and calculated the expected XUV signal based on corresponding ionization and coherent length $L_{coh}$ using (2) (full blue curve of Fig. 2b.). This XUV signal is well below the possible maximum. If we modify the medium by pre-ionizing it by 8%, the expected signal for the same laser pulse enhances significantly (red curve of Fig. 2b) and moves into the region of ideal absorption limited phase-matched generation, where the signal of 17$^{\text{th}}$ harmonic is 8 times stronger than its signal without the pre-ionization.

## IV. EXPERIMENTAL RESULTS

Experimental verification was performed at the High Harmonic Beamline at the ELI Beamlines facility [27] employing a Ti:Sapphire laser (Legend Elite Duo) with up to 40 fs pulses with central wavelength of 798 nm and energy of 10 mJ focused by a spherical mirror with the focal length of 5 m to drive the HHG. A 15 mm long glass capillary with a 1 mm inner diameter filled with Kr or Ar and hollow electrodes on both sides was placed in the laser focus for generation of a pre-ionized medium (see Fig 3). Note that the inner diameter of such a short capillary was too large for the laser beam to be guided, so its propagation can be considered as free-space propagation in contrast to the work reported in [16] and [17] where the laser was guided in a narrower capillary with a highly ionized medium.

The electrical discharge was generated by a RLC circuit (Fig. 3) [28], where energy is stored in a capacitor bank and then discharged through a 100 $\Omega$ resistor into the capillary. The current pulse had a 45 ns rising edge followed by an exponential decay with a characteristic time of 0.5 µs. The peak current was defined by charging voltages of 4 kV or 5 kV, corresponding to peak currents of 40 A or 50 A, respectively. Electrical capillary discharge was selected as the most suitable pre-ionization method due to its overall homogeneity (confirmed by MHD simulations [29]), its repeatability, and its simple construction allowing the repetition rate up to 1 kHz.

The discharge was synchronized with the laser so that the laser pulse interacted with the pre-ionized target 250 ns after the breakdown of the discharge. Generated harmonic radiation was characterized by an imaging flat-field XUV spectrometer [27]. The pressure inside the capillary and the peak current of the discharge were varied to control the phase-matching of HHG by varying the pre-ionization of the medium while keeping the driving laser parameters unchanged.

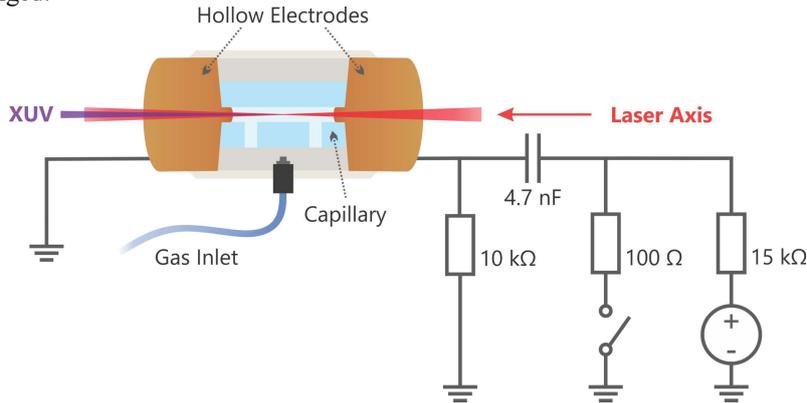

Fig. 3. Schematics of the gas-filled capillary with two hollow copper electrodes connected to the RLC circuit driving the capillary discharge to generate the pre-ionized medium for HHG.

The laser intensity during the experiment was set by aperturing the beam to reach harmonic cut-off frequency corresponding to 19$^{\text{rd}}$ harmonic order in krypton and 23$^{\text{rd}}$ harmonic order in Ar. The strength of the 17$^{\text{th}}$ harmonic signal measured in the experiment with Kr together with the estimate of degree of ionization from the inverse of equation (5) (Fig. 4a dashed purple line) indicate that for low gas pressures (below 25 mbar), the ionization is too high for efficient HHG. For pressures higher than 25 mbar, the combined ionization from the discharge and the laser propagation falls exactly into the phase-matching region, leading to optimum phase-matched generation. In particular, the highest XUV signal of the 17$^{\text{th}}$ harmonic is seen at 35 mbar of krypton for the 17$^{\text{th}}$ harmonic generated in krypton.

Figure 4b summarizes the enhancement factor for the H15, H17, and H19 in krypton with 40 A current discharge as a function of pressure. Similarly, Fig. 4c represents the measurements with 50 A discharge in argon for the H17, H19, and H21. For krypton, the ideal phase-matching is achieved when the total ionization degree is 9.6%, 8.7%, and 8.0% for the presented orders. Similarly, ideal ionization values for these harmonics generated in argon are 7.3%, 6.3%, and 5.6%, respectively. The highest measured XUV signal enhancement in argon (Fig. 4c) is by a factor of 5, compared to a factor of 6 observed in krypton (Fig. 4b). The optimum pre-ionization is seen for all three harmonic orders at 35 mbar and 40 A current pulse in krypton and 40 mbar and 50 A current pulse in argon, demonstrating the improved phase-matching conditions for a broad spectral range. Such a broadband effect is expected because the intervals of degree of ionization for close-by harmonics do not change significantly.

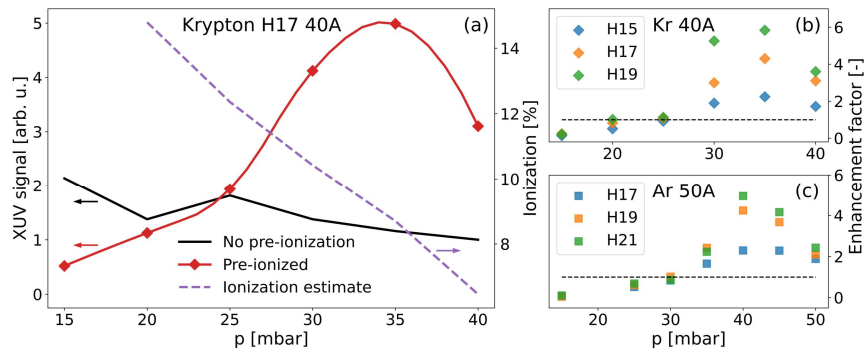

fig. 4. (a) Experimentally measured intensity of the 17$^{th}$ harmonic order as a function of pressure for HHG in Kr without pre-ionization (black curve) and in pre-ionized medium with discharge peak currents of 40 A (red diamonds). Red line (spline) is added to guide the eye. Estimate of degree of ionization in the capillary from the inverse of Eq. (5) with 35 mbar considered as maximum of the signal (purple dashed curve). (b) Enhancement factor for H15 (blue), H17 (orange), and H19 (green diamonds) in krypton with 40 A peak current. (c) Enhancement factor for H17 (blue), H19 (orange), and H21 (green squares) in argon with 50 A peak current. Each experimental point corresponds to the strongest shot from a set of ten laser shots.

## V. NUMERICAL SIMULATIONS

To support the analytical model and experimental data we performed numerical simulations containing both microscopic and macroscopic aspects of HHG. First, we computed the propagation of the driving laser in the medium by the cylindrically symmetric unidirectional solver [30]. This model intrinsically contains the dispersion of the driving fields by neutrals, plasma generation, and propagation in the plasma and all geometrical effects as well. Once we calculate both plasma and intensity profiles in space and time (on-axis values of the intensity are shown in Fig. 1a), we can study the phase-matching conditions.

To estimate the coherence length in the whole medium, we add the microscopic dipole phase obtained by the Saddle point approximation as explained in [31,32] and linear dispersion for a given harmonic order (see Supplement for more details). The spatial profile of coherence length of the 17$^{th}$ harmonic generated in Kr at a gas pressure of 35 mbar and peak laser intensity of $0.9 \times 10^{14}$ W/cm$^2$ is shown in Fig. 5a (no pre-ionization) and Fig. 5b (with pre-ionization, note the different color scale). As seen in those figures, the coherence length exceeds 30 mm i.e., double the length of the medium, in a large area for the 8% pre-ionized case. On the contrary, without the pre-ionization, it stays always below 1.5 mm in the region II (second half of the medium). Therefore, the XUV is efficiently generated, and the signal is added constructively from the extended region only in the pre-ionized case.

These simulations reveal as well that the major contribution to the phase-mismatch is given by the susceptibilities of the medium (neutrals and plasma). Other effects, such as the geometrical phase of the laser beam and the intensity-dependent phase of the harmonic dipole become significant only in the optimum pre-ionization case, where the indices of refraction of the laser and the harmonic field become similar (see Supplement for details). This fact justifies the application of the simple analytical model (Eq. 2), which is a plane-wave approximation of the signal built-up, in region II.

The on-axis coherence length at the end of the 15 mm-long krypton medium pre-ionized to 8% obtained from full numerical laser propagation code with various pressures and input laser intensities is shown in Fig 5c. Let us recall also Fig 1c, where the full on-axis profile of the associated phase mismatch $|\Delta k|$ is shown for several points of Fig. 5c. As seen in the figure, the generation with this level of pre-ionization is optimum for a broad range of pressures and driving laser intensities and demonstrates the robustness of our generation scheme.

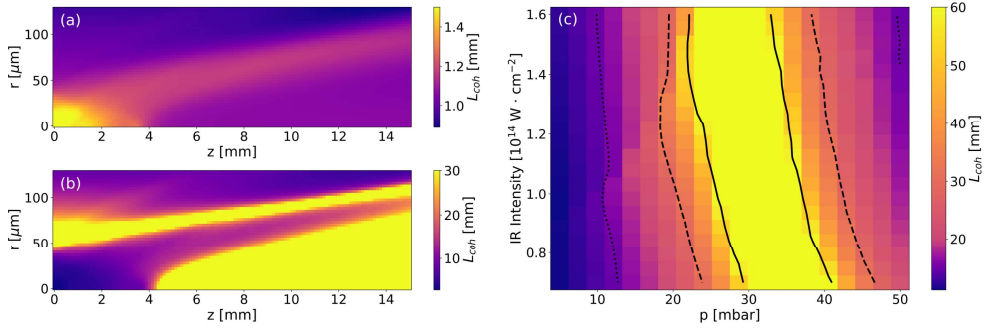

Fig. 5. Spatial distribution of the coherence length $L_{coh}$ at the peak of the laser pulse for the 17$^{th}$ harmonic generated in 35 mbar of Kr with vacuum intensity of $0.9 \cdot 10^{14}$ W/cm$^2$ and vacuum focus positioned at the center of the medium. Pre-ionization of the medium was set to 0 % (a) and 8 % (b). (c) The on-axis coherence lengths after 15 mm propagation at the peak of the laser pulse as a function of the input-laser peak intensity and medium pressure. $L_{coh} = L_{med}$ (dotted line), $L_{coh} = 2L_{med}$ (dashed line), and $L_{coh} = 4L_{med}$ (full line).

Furthermore, we performed a set of numerical simulations of the whole process. The level of pre-ionization at various pressures was calculated by one-dimensional dissipative magnetohydrodynamic (MHD) code utilizing a two-temperature model (ion and electron) and following dissipative processes: electron thermal conductivities, Joule heating, Nernst and Ettinghausen effects, radiation losses, and viscosity [33]. In the next step, we have used a full numerical model describing the HHG experiment also including the exact quantum dipoles from 1D time-dependent Schrödinger equation [34] and numerical XUV propagation to the far-field analogical to [35]. The resulting ionization and XUV signal as a function of the pressure is in Fig. 6a.

The simulated XUV signal as a function of pressure (Fig. 6a) shows the significant signal enhancement due to pre-ionization and agrees with the one-dimensional model (Eq. 1) and experimental observation (Fig. 4a). The highest amplification occurs around 8 % of ionization, which agrees with the degree of ionization estimated from the experiment using an independent simplified model based on Eq. 5. There is a small shift of the maximum of the simulated signal in pressure as compared to the experimental results, caused by possible underestimation of degree of ionization in the MHD simulations of the Kr discharge, which were used as the input for the complex simulations. The simulations also show that ionization by the laser field (in region II at the end of the medium) is small compared to the pre-ionization set by the discharge and alone cannot reach an optimal level of ionization.

Fig. 6b compares the simulated and experimentally measured XUV spectra for Kr at a pressure of 35 mbar and the value of pre-ionization set to optimal 8 %. We found a good agreement between theory and experiment in the whole spectral region studied. During the experiment we reach an enhancement of the XUV signal by factor of 6. The simulated spectrum shows an increase of a factor of up to 8, which is in excellent agreement with the value from the analytical model shown in Fig. 2c.

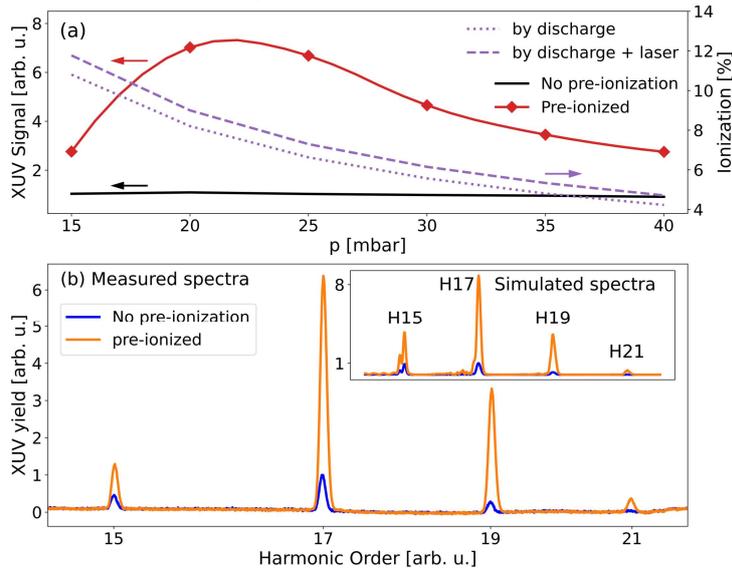

Fig. 6. (a) The simulated XUV signal and ionization level as a function of pressure. Full black line shows the signal without pre-ionization. XUV signal with pre-ionization is denoted by tehred diamonds. Red line (spline) is added to guide the eye. The dotted line shows the pre-ionization of the medium obtained from MHD simulations. The dashed line shows the estimate of the ionization on axis at the peak of the pulse in time and at the end of the medium. (b) Harmonic spectrum at 35 mbar in Kr comparing experimental data of optimal spectra (40 A peak current) with numerically computed spectrum for 8% pre-ionization.

The pre-ionization levels considered here are too low to affect the optical field ionization rates. Therefore, the propagation of the IR laser pulse is not affected by the pre-ionization of the medium and its wavefront, as well as the wavefront of the harmonic beam, remain unchanged. A comparison of far-field intensity patterns of the simulated harmonic beams generated at various pre-ionization levels (not shown here) confirmed their shape-independence of the medium ionization degree considering low ionization levels relevant to our study.

## VI. DISCUSSION & CONCLUSION

We have shown experimentally and derived theoretically that a few percent of a medium pre-ionization leads to significant improvement of the phase matching and, consequently, an increase of the generated XUV signal. We have employed a weak capillary discharge to achieve a controllable pre-ionization of the generating medium and reached up to six-fold enhancement of the XUV signal.

To establish the pre-ionization, other methods could also be used, such as radio-frequency discharge [36], a sequence of ionizing laser pulses [37], or mixing gas with low ionization potential acting as an electron donor on the rising edge of the laser pulse [15]. As long as the medium remains homogeneous, the weak uniform pre-ionization of the medium does not affect the propagation of the laser. Therefore, it does not affect the wavefront of the generated XUV beam, the divergence of the harmonic beam, and the position of the XUV source.

In summary, we have demonstrated a new way of controlling the phase-matching of the HHG by pre-ionizing the generating medium. It increases the HHG conversion efficiency in the absorption-limited generation in long media. As the pre-ionization does not affect any other aspect of the generation, this mechanism turns out to be an independent tuning parameter of the experiment, paving the way for enabling photon-hungry applications with compact HHG sources.

Moreover, the broadband nature of the method ensures that the temporal structure of the XUV pulse remains unchanged, which opens a way for its implementation within the generation of single attosecond pulses. Another possible application might be employment of our method in HHG with picosecond laser pulses that suffer from fast over-ionization of the medium resulting in strong defocusing and overall low conversion efficiency. Interaction of lower intensity ps pulses within a properly pre-ionized medium could significantly improve the generation of very spectrally narrow harmonics.

## ACKNOWLEDGMENTS

We are grateful to Michael Greco for manuscript revision. This research was supported by the project ADONIS (CZ.02.1.01/0.0/0.0/16_019/0000789) and CAAS (CZ.02.1.01/0.0/0.0/16-019/0000778) from European Regional Development Fund and by Ministry of Education, Youth and Sports within targeted support of Large infrastructures (project LM2018141). S. Skupin acknowledges "Qatar National Research Fund (NPRP 12S-0205-190047)" and "Grand Équipement National De Calcul Intensif (GENCI A0100507594)"

**Disclosures.** The authors declare no conflicts of interest

**Data availability.** Data underlying the results presented in this paper are not publicly available at this time but may be obtained from the authors upon reasonable request.

## REFERENCES

1  M. Th. Hassan, T. T. Luu, A. Moulet, O. Raskazovskaya, P. Zhokhov, M. Garg, N. Karpowicz, A. M. Zheltikov, V. Pervak, F. Krausz, and E. Goulielmakis, *Optical attosecond pulses and tracking the nonlinear response of bound electrons*, Nature **530**, 66–70, (2016).

2  G. M. Rossi, R. E. Mainz, Y. Yang, F. Scheiba, M. A. Silva-Toledo, S-H. Chia, P. D. Keathley, S. Fang, O. D. Mücke, C. Manzoni, G. Cerullo, G. Cirmi. and F. X. Kärtner, *Sub-cycle millijoule-level parametric waveform synthesizer for attosecond science*, Nature Photonics **14**, 629-635, (2020)

3  S. Schoun, A. Camper, P. Salieres, R. Lucchese, P. Agostini, and L. Dimauro, *Precise Access to the Molecular-Frame Complex Recombination Dipole through High-Harmonic Spectroscopy*, Phys. Rev. Lett. **118**, 10.1103, (2017).

4  P. B. Corkum, *Plasma perspective on strong field multiphoton ionization*, Phys. Rev. Lett. **71**, 1994 (1993)

5  M. B. Gaarde, J. Tate, and K. Schafer, *Macroscopic aspects of attosecond pulse generation*, Journal of Physics B: Atomic, Molecular and Optical Physics. **41**. 132001. 10.1088/0953-4075/41/13/132001 (2008).

6  T. Popmintchev, M-Ch. Chen, A. Bahabad, M. Gerrity, P. Sidorenko, O. Cohen, I. P. Christov, M. M. Murnane, and H. C. Kapteyn, *Phase matching of high harmonic generation in the soft and hard X-ray* *regions of the spectrum*, Proceedings of the National Academy of Sciences, 106 (26) 10516-10521 (2009).

7. R. Budriūnas, T. Stanislauskas, J. Adamonis, A. Aleknavičius, G. Veitas, D. Gadonas, S. Balickas, A. Michailovas, and A. Varanavičius, *53 W average power CEP-stabilized OPCPA system delivering 5.5 TW few cycle pulses at 1 kHz repetition rate*, Opt. Express 25, 5797-5806 (2017).
8. P. Bakule, R. Antipenkov, J. Novák, F. Batysta, R. Boge, J. T. Green, Z. Hubka, M. Greco, L. Indra, A. Špaček, J. A. Naylon, K. Majer, P. Mazůrek, E. Erdman, V. Šobr, B. Tykalewicz, T. Mazanec, P. Strkula, and B. Rus, *Readiness of L1 ALLEGRA Laser System for User Operation at ELI Beamlines*, in OSA High-brightness Sources and Light-driven Interactions Congress 2020 (EUVXRAY, HILAS, MICS)
9. C. M. Heyl, H. Coudert-Alteirac, M. Miranda, M. Louisy, K. Kovacs, V. Tosa, E. Balogh, K. Varjú, A. L'Huillier, A. Couairon, and C. L. Arnold, *Scale-invariant nonlinear optics in gases*, Optica **3**, 75-81 (2016)
10. P. Rudawski, C. M. Heyl, F. Brizuela, J. Schwenke, A. Persson, E. Mansten, R. Rakowski, L. Rading, F. Campi, B. Kim, P. Johnsson, and A. L'Huillier, *A high-flux high-order harmonic source*, The Review of scientific instruments. 84. 073103. 10.1063/1.4812266 (2013).
11. E. Takahashi, Y. Nabekawa, and K. Midorikawa, *Generation of 10-μJ coherent extreme-ultraviolet light by use of high-order harmonics,* Opt. Lett. 27, 1920-1922 (2002).
12. L. Roos, E. Constant, E. Mével, P. Balcou, D. Descamps, M. Gaarde, A. Valette, R. Haroutunian, and A. L'Huillier, *Controlling phase matching of high-order harmonic generation by manipulating the fundamental field*, Phys. Rev. A. 60. 10.1103/PhysRevA.60.5010 (1999).
13. P. Balcou, R. Haroutunian, S. Sebban, G. Grillon, A. Rousse, G. Mullot, J. Chambaret, G. Rey, A. Antonetti, D. Hulin, L. Roos, D. Descamps, M. Gaarde, A. L'Huillier, E. Constant, E. Mevel, D. von der Linde, A. Orisch, A. Tarasevitch, and W. Theobald, *High-order-harmonic generation: Towards laser-induced phase-matching control and relativistic effects*, Applied Physics B: Lasers and Optics. 74. 10.1007/s003400200797 (2002).
14. E. Constant, A. Dubrouil, O. Hort, S. Petit, D. Descamps, and E. Mével, *Spatial shaping of intense femtosecond beams for the generation of high-energy attosecond pulses*, Journal of Physics B: Atomic, Molecular and Optical Physics, **45**, 074018, 10.1088/0953-4075/45/7/074018 (2012).
15. L. Wang, W. Zhu, H. Li, and Y. Zhang, *Spectrum modification of high-order harmonic generation in a gas mixture of Ar and Kr*, J. Opt. Soc. Am. B 35, A39-A44 (2018).
16. B. A. Reagan, T. Popmintchev, M. E. Grisham, D. M. Gaudiosi, M. Berrill, O. Cohen, B. C. Walker, M. M. Murnane, J. J. Rocca, and H. C. Kapteyn, *Enhanced high-order harmonic generation from Xe, Kr, and Ar in a capillary discharge*, Phys. Rev. A. **76**, 013816 (2007).
17. D. M. Gaudiosi, B. Reagan, T. Popmintchev, M. Grisham, M. Berrill, O. Cohen, B. C. Walker, M. M. Murnane, H. C. Kapteyn, J. J. Rocca, *High-Order Harmonic Generation from Ions in a Capillary Discharge*, Physical Review Letters. **96**, 203001 (2006).
18. K. Cassou, S. Daboussi, O. Hort, O. Guilbaud, D. Descamps, S. Petit, E. Mével, E. Constant, and S. Kazamias, *Enhanced high harmonic generation driven by high-intensity laser in argon gas-filled hollow core waveguide*, Opt. Lett. **39**, 3770-3773 (2014).
19. A. Becker, N. Aközbek, K. Vijayalakshmi, E. Oral, C.M. Bowden & S.L. Chin, *Intensity clamping and re-focusing of intense femtosecond laser pulses in nitrogen molecular gas*, Appl Phys B 73, 287–290 (2001).
20. H. R. Lange, A. Chiron, J.-F. Ripoche, A. Mysyrowicz, P. Breger, and P. Agostini, *High-Order Harmonic Generation and Quasiphase Matching in Xenon Using Self-Guided Femtosecond Pulses*, Phys. Rev. Lett. 81, 1611 (1998).
21. D. E. Rivas, B. Major, M. Weidman, W. Helml, G. Marcus, R. Kienberger, D. Charalambidis, P. Tzallas, E. Balogh, K. Kovács, V. Tosa, B. Bergues, K. Varjú, and L. Veisz, *Propagation-enhanced generation of intense high-harmonic continua in the 100-eV spectral region*, Optica 5, 1283-1289 (2018).
22. T. Brabec and F. Krausz, *Intense few-cycle laser fields: Frontiers of nonlinear optics*, Rev. Mod. Phys. 72, p. 545 (2000).
23. E. Constant, D. Garzella, P. Breger, E. Mével, Ch. Dorrer, C. Le Blanc, F. Salin, and P. Agostini, *Optimizing High Harmonic Generation in Absorbing Gases: Model and Experiment*, Phys. Rev. Lett. 82, 1668, (1999).
24. A. Bideau-Mehu, Y. Guern, R. Abjean, and A. Johannin-Gilles, *Measurement of refractive indices of neon, argon, krypton and xenon in the 253.7-140.4 nm wavelength range. Dispersion relations and estimated oscillator strengths of the resonance lines*, J. Quant. Spectrosc. Rad. Transfer 25, 395-402 (1981).

25  B.L. Henke, E.M. Gullikson, and J.C. Davis, *X-ray interactions: photoabsorption, scattering, transmission, and reflection at E=50-30000 eV, Z=1-92*, Atomic Data and Nuclear Data Tables Vol. **54 (no.2)**, 181-342 (1993).
26  C.T. Chantler, K. Olsen, R.A. Dragoset, J. Chang, A.R. Kishore, S.A. Kotochigova, and D.S. Zucker, *Detailed Tabulation of Atomic Form Factors, Photoelectric Absorption and Scattering Cross Section, and Mass Attenuation Coefficients for Z = 1-92 from E = 1-10 eV to E = 0.4-1.0 MeV*, NIST, Physical Measurement Laboratory (2001).
27  O. Hort, M. Albrecht, V. E. Nefedova, O. Finke, D. D. Mai, S. Reyné, F. Giambruno, F. Frassetto, L. Poletto, J. Andreasson, J. Gautier, S. Sebban, and J. Nejdl, *High-flux source of coherent XUV pulses for user applications*, Opt. Express 27, 8871-8883 (2019).
28  M.F. Nawaz, M. Nevrkla, A. Jancarek, A. Torrisi, T. Parkman, J. Turnova, L. Stolcova, M. Vrbova, J. Limpouch, L. Pina, and P. Wachulak, *Table-top water-window soft X-ray microscope using a Z-pinching capillary discharge source*, JINST **11** P07002, (2016).
29  G. A. Bagdasarov, P. V. Sasorov, V. A. Gasilov, A. S. Boldarev, O. G. Olkhovskaya, C. Benedetti, S. S. Bulanov, A. Gonsalves, H.-S. Mao, C. B. Schroeder, J. van Tilborg, E. Esarey, W. P. Leemans, T. Levato, D. Margarone, and G. Korn, *Laser beam coupling with capillary discharge plasma for laser wakefield acceleration applications*, Physics of Plasmas **24**, 083109 (2017).
30  L. Bergé, S. Skupin, R. Nuter, J. Kasparian, and J-P Wolf, *Ultrashort filaments of light in weakly ionized, optically transparent media*, Rep. Prog. Phys. **70** 1633, (2007).
31  M. B. Gaarde, F. Salin, E. Constant, Ph. Balcou, K. J. Schafer, K. C. Kulander, and A. L'Huillier, *Spatiotemporal separation of high harmonic radiation into two quantum path components*, Phys. Rev. A **59**, 1367 (1999).
32  M. B. Gaarde, A. L'Huillier, and M. Lewenstein, *Theory of high-order sum and difference frequency mixing in a strong bichromatic laser field*, Phys. Rev. A **54**, 4236 (1996).
33  N. A. Bobrova, A. A. Esaulov, J.-I. Sakai, P. V. Sasorov, D. J. Spence, A. Butler, S. M. Hooker, and S. V. Bulanov, *Simulations of a hydrogen-filled capillary discharge waveguide*, Physical Review. E, **65**. 016407, (2002).
34  F. Catoire and H. Bachau, *Above-Threshold Ionization of Quasiperiodic Structures by Low-Frequency Laser Fields*, Phys. Rev. Lett. **115**, 163602, 10.1103/PhysRevLett.115.163602, (2015).
35  F. Catoire, A. Ferré, O. Hort, A. Dubrouil, L. Quintard, D. Descamps, S. Petit, F. Burgy, E. Mével, Y. Mairesse, and E. Constant, *Complex structure of spatially resolved high-order-harmonic spectra*, Phys. Rev. A **94**, 063401, (2016).
36  P. Chabert, and N. Braithwaite, *Physics of Radio Frequency plasmas*, (Cambridge University Press, 2011).
37  Y.-F. Xiao, and H.-H. Chu, *Efficient generation of extended plasma waveguides with the axicon ignitor-heater scheme*, Physics of Plasmas **11**, L21, (2004).

# Phase-matched high-order harmonic generation in pre-ionized noble gases: supplemental document

This Supplement of "Phase-matched high-order harmonic generation in pre-ionized noble gases": 1) explains the methodology of the analysis of the numerical data to retrieve the quantities related to the phase of the field, 2) provides more detailed analysis of the various physical models from simple analytical formulas up to a fully numerical treatment.

## 1. PROCEDURES TO ANALYSE THE NUMERICAL SIMULATIONS

Here, we present the procedure to retrieve the phase of the propagating laser field and the spatial distribution of the coherence length used in Figs. 1 (b), 1 (c) and 5 of the main text. To get an insight, let us start with an analytical model of a complex on-axis field, i.e. we consider an additional geometrical phase, $\phi_{\text{geom.}}$. The field is written as

$$E_{\text{IR}}(z,t) \propto e^{-\mathbf{i}\left(\omega_0 t - \frac{\omega_0}{c}\int_0^z n(z',t)\,dz' + \phi_{\text{geom.}}\right)} = e^{-\mathbf{i}(\omega_0 t - \Phi(z,t))}. \tag{S1}$$

This field induces harmonic response at $z_1$ written as

$$E_{\text{XUV}}^{(1)}(z_1,t) \propto e^{-\mathbf{i}\left(q(\omega_0 t - \Phi(z_1,t)) + \phi_{\text{atom},q}(z_1,t)\right)}, \tag{S2}$$

where the first part of the phase, $q(\omega_0 t - \Phi(z_1,t))$, is imprinted by the driving field, and $\phi_{\text{atom},q}(z_1,t)$ is the phase related to the microscopic generation process of the given harmonic order $q$. Let us propagate this field infinitesimally to $z_2 = z_1 + \Delta z$, which means only add $q\omega_0 n_q \Delta z/c$ to the phase. $n_q$ is the refractive index of the harmonic field. If we compare the field generated at $z_1$ and propagated to $z_2$ with the field generated at $z_2$; we have

$$E_q^{(1)}(z_2,t) \propto e^{-\mathbf{i}\left(q(\omega_0 t - \Phi(z_1,t)) + \phi_{\text{atom},q}(z_1,t)\right) + \mathbf{i}\frac{q\omega_0}{c}n_q \Delta z}, \tag{S3a}$$

$$E_q^{(2)}(z_2,t) \propto e^{-\mathbf{i}\left(q(\omega_0 t - \Phi(z_2,t)) + \phi_{\text{atom},q}(z_2,t)\right)}, \tag{S3b}$$

respectively. We subtract the phases

$$q\left(\Phi(z_2,t) - \Phi(z_1,t)\right) - \left(\phi_{\text{atom},q}(z_2,t) - \phi_{\text{atom},q}(z_1,t)\right) + \frac{q\omega_0}{c}n_q \Delta z \approx$$

$$\approx \left(\frac{q\omega_0}{c}n_q - q\frac{\partial \Phi(z_1,t)}{\partial z} - \frac{\partial \phi_{\text{atom},q}(z_1,t)}{\partial z}\right)\Delta z = \Delta k_q(z_1,t)\Delta z, \tag{S4}$$

we Taylorized the infinitesimal difference of the phases. It provided the phase mismatch, which we denoted $\Delta k_q(z_1,t)$. Because we are within an infinitesimal interval, we can omit the subscript in $z_1$, the result is a local quantity

$$\Delta k_q(z,t) = \frac{q\omega_0}{c}n_q - q\frac{\partial \Phi(z,t)}{\partial z} - \frac{\partial \phi_{\text{atom},q}(z,t)}{\partial z} \tag{S5}$$

with the associated coherence length $L_{\text{coh},q} = |\pi/\Delta k_q|$. An important property is that the result is a function of $t$. This is justified because the time evolution of the phases is dominant compared to the frequency, which is a natural variable of the refractive-index. We thus evaluate all the quantities at the central frequency of the driver or at the harmonic orders (the aspect of time in the refractive index is discussed in details in [1]).

To evaluate the last term in Eq. (S5), we need the atomic phase $\phi_{\text{atom}}$. We use the Saddle-point calculation [2, 3],

$$\phi_q(I) = \phi_q(I_a) + \frac{\partial \phi_q(I_a)}{\partial I}(I - I_a) + \cdots = \phi_q(I_a) - \alpha_q(I_a)(I - I_a) + \cdots \tag{S6}$$

in the vicinity of the driver-field intensity $I_a$. $\alpha_q(I_a)$ is introduced to match the usual expression as used in [4]. The referred works [2–4] show that it is well-approximated by taking only the linear term in the expansion. Using only this term, we obtain

$$\frac{\partial \phi_{\text{atom}}(z,t)}{\partial z} = -\alpha(I(z,t))\frac{\partial I(z,t)}{\partial z}. \tag{S7}$$

It means this term is locally proportional to the gradient of the intensity.

An advantage of Eq. (S5) is that it can be used for any complex electric field of the form $E_{\text{IR}} \propto \exp(-\mathbf{i}(\omega_0 t - \Phi(\vec{r},t)))$, even if the decomposition assumed in Eq. (S1) is not possible. The phase is retrieved after removing fast oscillations as

$$\Phi(\vec{r},t) = \text{Arg}\left(e^{\mathbf{i}\omega_0 t} E(\vec{r},t)\right). \tag{S8}$$

This formula is used to process the numerically computed laser fields $E(\vec{r},t)$ obtained from the laser-propagation code.

Using the full profile of the laser pulse, $\Delta k_q$ becomes fully spatio-temporal quantity in $(t,\vec{r})$. We will need to choose some cuts to visualise it. To analyse the spatial dependence, we fix time in the peak of the pulse, where the maximal HHG production is expected; this procedure agrees with the one used in [5]. Note, that the homogeneous pre-ionization is only a constant shift in $\Delta k_q(z,t)$, the choice of the reference time is then fully consistent across various initial conditions.

Finally, we return to the physical interpretation of Eq. (S5). It is rewritten with the help of the relation between indexes of refraction as $n \approx 1 + \chi_{\text{disp.}}/2 + \chi_{\text{plasma}}/2$ as

$$\Delta k_q = \underbrace{\frac{q\omega_0}{2c}\left(\chi_{\text{disp.},q} - \chi_{\text{disp.,IR}}\right)}_{\Delta k_{\text{disp.}}} - \underbrace{\frac{q\omega_0}{2c}\chi_{\text{plasma,IR}}}_{\Delta k_{\text{plasma}}} - \underbrace{q\frac{\partial \phi_{\text{geom.}}}{\partial z}}_{\Delta k_{\text{geom.}}} - \underbrace{\frac{\partial \phi_{\text{atom},q}}{\partial z}}_{\Delta k_{\text{atom}}}. \tag{S9}$$

The plasma effect is not taken into account for the XUV-field as the induced phase is negligible in the spectral range under consideration. This is exactly the usual formula used to study phase-matching from [6]. We stress here that all the effects are present for our numerical field. The stable intensity region of the interest means small gradients of intensity and geometrical phase. The role of $\Delta k_{\text{geom.}}$ and $k_{\text{atom}}$ is thus minor in this region. The details of this assumption follow in section 2 by comparing the full-numerical and analytical models.

## 2. BENCHMARKING MODELS: PHASE MISMATCH FROM IONISATION RATIO $\eta$

We have used the analytical model of the mismatch, Eq. (3) in the main manuscript, for a basic picture of the problem; its only variable is the ionization degree $\eta$, which is a much simpler quantity than the phase Eq. (S8). Additionally, we have introduced an *ab initio* numerical method to compute the same quantity in the previous paragraph Eq. (S5). Natural questions are: How do these two compare? Do the analytical and numerical models agree quantitatively? In other words, we now benchmark these models.

We start with the phase mismatches shown in Fig. 1(b,c), which were obtained from the numerical analysis Eq. (S5). Figure S1 compares these values with the analytical model, where only $\eta$ is taken from the numerical propagation. The two models agree well in the stable region without the pre-ionization (panels (a) and (b)). The scale is also correct with the 8 % pre-ionization, but the results differ more (panels (c) and (d)). This is not surprising, because the differences from other effects – the beam geometry and the microscopic phase of the dipole – are at play once the dominant effect of the ionization is removed. However, these differences remain within the range that is still favorable for phase-matching. In summary, these results confirm that the simple model describes well the dominant effect and may be used for an optimization procedure.

We compare also the maps of the coherence length shown in Fig. 5. The result is in Fig. S2. The overall scale is also correct and the models agree well in the most important region: near the axis in the stabilized region II. There are differences in the off-axis region ($r > 50$ $\mu$m), where intensity gradients make the atomic dipole phase more significant. Because $\eta$ is the only variable, the constant $L_{\text{coh}}$ for $r > 50$ $\mu$m in the right panels shows that there is no ionisation. It means that the dominant region for the harmonic generation is $r < 50$ $\mu$m. The non-trivial evolution of the phase outside this region for the numerical model is related to the transverse gradient of the field.

We recall that these maps are only snapshots in the peak of the pulse. We have also investigated the change with different time near by the peak of the field envelope. The main trend, however, remained stable. Since the non-linear propagation is a complex process, a snapshot in time gives only a strong indication that an optimal $L_{\text{coh}}$ would lead to a large gain. The benchmark of this is the fully numerical model.

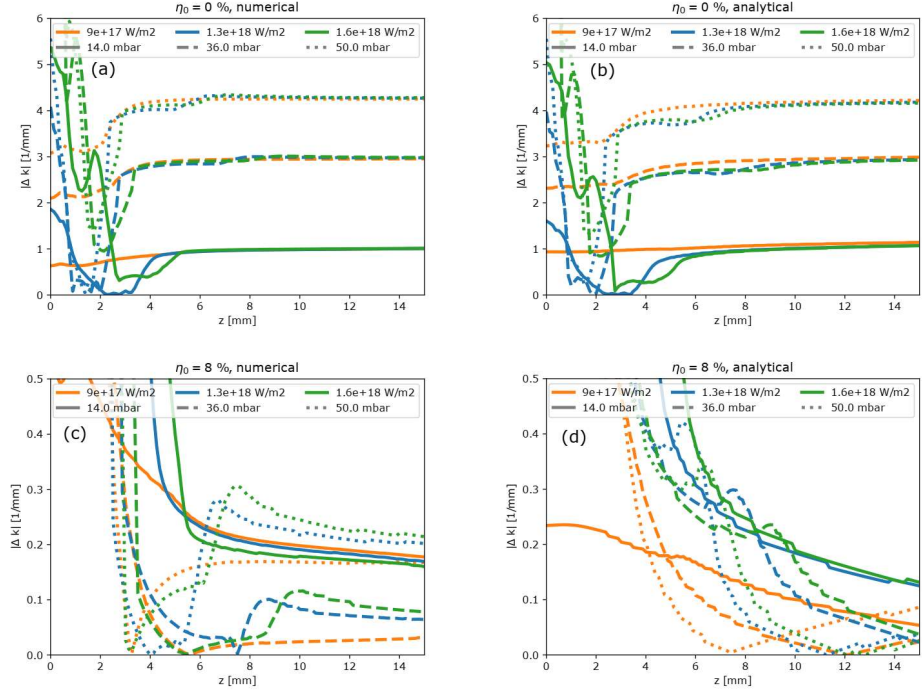

**Fig. S1.** The comparison of the two models for computing $\Delta k$ (numerical based on Eq. (S5) and analytical based on Eq. (3)). Left panels show the results of the complete calculation already discussed in Fig. 1(b,c), the right panels are their counterparts computed from the analytic approach using ionization ratios at the peak of the pulse obtained from the numerical laser propagation.

In conclusion, we have proven that the analytical model works well to describe the dominant effect – the ionization – in the spatial region relevant for HHG. Next, it is also a good benchmark of the code: Although the initial electric field and plasma density come both from the same propagation code, the treatment of the phase is then completely *independent*. In the first case, the phase mismatch is obtained from the numerical differentiation of the phase of the numerical field. In the second case, it is only the degree of ionization $\eta$ as a result of the propagation code used in the analytical model. Finally, all these results are consistent and agree with the fully numerical model from the main manuscript. It proves the ability of the analytical model to be predictive for the physics investigated.

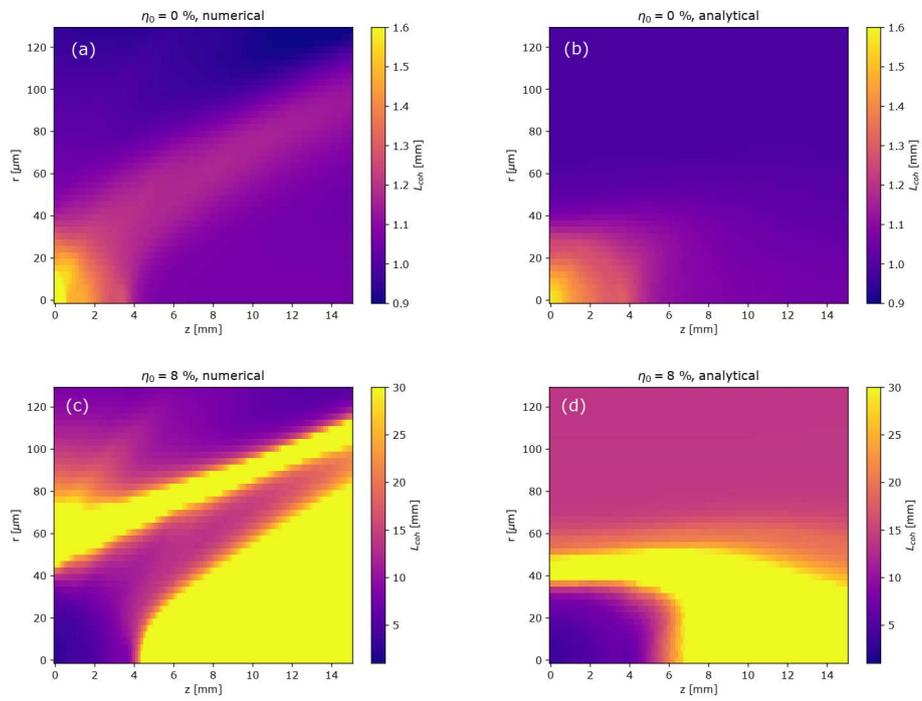

**Fig. S2.** The comparison of the two models for computing $L_{\mathrm{coh}}$. Left panels show the results presented in Fig. 5, the right panels are their counterparts computed from the analytic approach, similarly to Fig. S1.

**REFERENCES**

1. M. A. Khokhlova and V. V. Strelkov, "Highly efficient XUV generation via high-order frequency mixing," New J. Phys. **22**, 093030 (2020).
2. M. B. Gaarde, A. L'Huillier, and M. Lewenstein, "Theory of high-order sum and difference frequency mixing in a strong bichromatic laser field," Phys. Rev. A **54**, 4236–4248 (1996).
3. M. B. Gaarde, F. Salin, E. Constant, P. Balcou, K. J. Schafer, K. C. Kulander, and A. L'Huillier, "Spatiotemporal separation of high harmonic radiation into two quantum path components," Phys. Rev. A **59**, 1367–1373 (1999).
4. F. Catoire, A. Ferré, O. Hort, A. Dubrouil, L. Quintard, D. Descamps, S. Petit, F. Burgy, E. Mével, Y. Mairesse, and E. Constant, "Complex structure of spatially resolved high-order-harmonic spectra," Phys. Rev. A **94**, 063401 (2016).
5. D. E. Rivas, B. Major, M. Weidman, W. Helml, G. Marcus, R. Kienberger, D. Charalambidis, P. Tzallas, E. Balogh, K. Kovács, V. Tosa, B. Bergues, K. Varjú, and L. Veisz, "Propagation-enhanced generation of intense high-harmonic continua in the 100-ev spectral region," Optica **5**, 1283–1289 (2018).
6. C. Delfin, C. Altucci, F. D. Filippo, C. de Lisio, M. B. Gaarde, A. L'Huillier, L. Roos, and C.-G. Wahlström, "Influence of the medium length on high-order harmonic generation," J. Phys. B: At. Mol. Opt. Phys. **32**, 5397–5409 (1999).